\documentclass[twocolumn]{aastex6}
\usepackage{times}
\usepackage{amsmath}
\usepackage{textcomp}
\usepackage{amssymb}

\newcommand{\lum}{erg\,s$^{-1}$}
\newcommand{\fermi}{{\it Fermi}}

\newcommand{\phflux}{\mbox{${\rm \, ph \,\, cm^{-2} \, s^{-1}}$}}

\newcommand{\gm}{$\gamma$}

\shorttitle{\gm-NLSy1 galaxies}

\begin{document}

\title{Gamma-ray Emitting Narrow Line Seyfert 1 Galaxies in The Sloan Digital Sky Survey}
\author{
Vaidehi S. Paliya\altaffilmark{1}, M. Ajello\altaffilmark{1}, S. Rakshit\altaffilmark{2,3}, A. Mandal\altaffilmark{2}, C. S. Stalin\altaffilmark{2}, A. Kaur\altaffilmark{1}, and D. Hartmann\altaffilmark{1}}
\altaffiltext{1}{Department of Physics and Astronomy, Clemson University, Kinard Lab of Physics, Clemson, SC 29634-0978, USA}
\altaffiltext{2}{Indian Institute of Astrophysics, Block II, Koramangala, Bangalore, India, 560034}
\altaffiltext{3}{Astronomy Program, Department of Physics and Astronomy, Seoul National University, Seoul 151-742, Republic of Korea}
\email{vpaliya@g.clemson.edu}
\begin{abstract}
The detection of significant \gm-ray emission from radio-loud narrow line Seyfert 1 (NLSy1s) galaxies enables us to study jets in environments different than those in blazars. However, due to the small number of known \gm-ray emitting NLSy1 (\gm-NLSy1) galaxies, a comprehensive study could not be performed. Here we report the first detection of significant \gm-ray emission from four active galactic nuclei (AGN), recently classified as NLSy1 from their Sloan Digital Sky Survey (SDSS) optical spectrum. Three flat spectrum radio quasars (FSRQs) present in the third Large Area Telescope AGN catalog (3LAC) are also found as \gm-NLSy1 galaxies. Comparing the \gm-ray properties of these objects with 3LAC blazars reveals their spectral shapes to be similar to FSRQs, however, with low \gm-ray luminosity ($\lesssim10^{46-47}$ \lum). In the Wide-field Infrared Survey Explorer color-color diagram, these objects occupy a region mainly populated by FSRQs. Using the H$_{\beta}$ emission line parameters, we find that on average \gm-NLSy1 have smaller black hole masses than FSRQs at similar redshifts. In the low-resolution SDSS image of one of the \gm-NLSy1 source, we find the evidence of an extended structure. We conclude by noting that overall many observational properties of \gm-NLSy1 sources are similar to FSRQs and therefore, these objects could be their low black hole mass counterparts, as predicted in the literature.

\end{abstract}


\section{Introduction} \label{sec:intro}
The \gm-ray emitting narrow line Seyfert 1 (\gm-NLSy1) galaxies are an enigmatic member of the radio-loud active galactic nuclei (RL-AGN) hosting powerful relativistic jets. In a broader context, the NLSy1 galaxies are defined from their optical spectral properties with narrow Balmer lines (FWHM H$_{\beta}<2000$ km s$^{-1}$), weak [O~{\sc iii}], and strong Fe~{\sc ii} emission lines \citep[][]{1985ApJ...297..166O,1989ApJ...342..224G}. Only a small fraction ($\sim5-7\%$) of the NLSy1 galaxy population is found to be RL \citep[e.g.,][]{2006AJ....132..531K} and, in general, these objects are known to exhibit an efficient accretion process, close to the Eddington limit, onto relatively low-mass black holes \citep[10$^{6-8} M_\sun$;][]{2004ApJ...606L..41G}. However, there are claims that black hole masses in RL-NLSy1 galaxies are similar to RL quasars \citep[e.g.,][]{2008ApJ...678..693M,2016MNRAS.458L..69B}. A few RL-NLSy1 galaxies are found to display blazar like physical characteristics such as the large amplitude flux and spectral variability, compact radio cores, flat radio spectra ($-0.5<\alpha<0.5, F_\nu\propto \nu^{\alpha}$), and high brightness temperatures, thus indicating the presence of closely aligned relativistic jets \citep[][]{2008ApJ...685..801Y}. The unambiguous confirmation, however, came with the detection of the significant \gm-ray emission from about half-a-dozen RL-NLSy1 galaxies by the \fermi-Large Area Telescope \citep[\fermi-LAT;][]{2009ApJ...707L.142A,2011nlsg.confE..24F,2015MNRAS.452..520D,2015MNRAS.454L..16Y}. 
Intense multi-wavelength follow-up studies have further provided evidences supporting their similarity with blazars, in particular with flat spectrum radio quasars \citep[FSRQs, e.g.,][]{2013MNRAS.428.2450P,2014ApJ...789..143P,2015A&A...575A..13F,2016ApJ...820...52P}. Interestingly, the fact that \gm-NLSy1s could be hosted by spiral galaxies \citep[e.g.,][]{2016ApJ...832..157K}, similar to other NLSy1 sources \citep[][]{2006AJ....132..321D}, makes them different from blazars which are known to be powered by massive black holes hosted by elliptical galaxies. Therefore, it is of great importance to find new \gm-NLSy1 galaxies to characterize the physical properties of relativistic jets at different accretion and mass scales and to also understand whether \gm-NLSy1s are a different population of the \gm-ray emitting sources or they are low luminosity-low black hole mass counterpart of the powerful FSRQs.

Until now, a majority of the studies of NLSy1 galaxies were primarily based on the sample of 2011 AGNs, classified as NLSy1s by \citet[][]{2006ApJS..166..128Z} using the Sloan Digital Sky Survey Data Release 3 (SDSS-DR3). Since the classification of an AGN being a NLSy1 or not is purely based on its optical spectral properties, in order to identify more \gm-NLSy1 galaxies, one has to first increase the population size of the known NLSy1s using the latest advancements in the field of the optical spectroscopy. Such a systematic effort has been done recently by \citet[][]{2017ApJS..229...39R} using the SDSS-DR12. They have classified a total of 11,101 AGNs as NLSy1s, thus increasing the sample size by a factor of $\sim$5. Taking the advantage of this enlarged sample, we search for new \gm-NLSy1 galaxies and this work we report 7 new $\gamma$-NLSy1s, including two candidate NLSy1 sources. Four of them are found as the result of a specific analysis of 8 years of LAT data and three of them are noted as misclassified FSRQs in the third LAT AGN catalog \citep[3LAC;][]{2015ApJ...810...14A}.

Throughout, we adopt Hubble constant $H_0=67.8$~km~s$^{-1}$~Mpc$^{-1}$, $\Omega_m = 0.308$, and $\Omega_\Lambda = 0.692$ \citep[][]{2016A&A...594A..13P}.

\section{Sample selection and Analysis} \label{sec:sample_red}
Since \gm-ray detected blazars, and also \gm-NLSy1s, are known to be strong radio emitters, we cross-match all 11,101 NLSy1 galaxies with the NRAO-VLA Sky Survey \citep[NVSS;][]{1998AJ....115.1693C} to find the NLSy1 sources detected at radio wavelengths. This is done by searching for a NVSS counterpart within 10 arcseconds of the SDSS position, which is the typical uncertainty in the radio position in the NVSS survey \citep[][]{1998AJ....115.1693C}. At this stage, we are left with 259 radio detected NLSy1 galaxies. We then cross-match this updated sample with 3LAC. This is carried out to identify the known \gm-ray emitting sources that are NLSy1 galaxies, originally classified as FSRQs or BL Lac objects in 3LAC \citep[see, e.g.,][for a similar finding]{2015MNRAS.454L..16Y}. In addition to the already known \gm-NLSy1s, i.e., 3FGL J0849.9+5108 ($z=0.58$), 3FGL J0948.8+0021 ($z=0.58$), 3FGL J1505.1+0326  ($z=0.41$), and 3FGL J1644.4+2632\footnote{Note that two other NLSy1s which are present in 3LAC, 3FGL J0325.2+3410 ($z=0.06$) and 3FGL J2007.8$-$4429 ($z=0.24$), are outside of the SDSS coverage area.} \citep[$z=0.15$;][]{2015MNRAS.452..520D}, we find three more NLSy1 galaxies, 3FGL J0937.7+5008 ($z=0.28$), 3FGL J1520.3+4209 ($z=0.49$), and 3FGL J2118.4+0013 ($z=0.46$). All of them were classified as FSRQs in 3LAC and therefore, following \citet[][]{2017ApJS..229...39R}, they can now be considered as \gm-NLSy1 galaxies. Note that 3FGL J0937.7+5008 exhibits a relatively weak Fe~{\sc ii} emission \citep[Fe~{\sc ii}/H$_\beta = 0.05$;][]{2017ApJS..229...39R} contrary to that typically seen in NLSy1 sources. However, recently \citet[][]{2016MNRAS.462.1256C} have argued that a strong Fe {\sc ii} emission may not be a characteristic feature of NLSy1 galaxies. A weak Fe~{\sc ii} emission has also been observed from another \gm-NLSy1 galaxy 3FGL J2007.8$-$4429 \citep[][]{2001ApJ...558..578O,2006MNRAS.370..245G}. Since all of the other multi-wavelength properties of 3FGL J0937.7+5008 are similar to the known \gm-NLSy1 objects, we include it in our sample, though it would be appropriate to call this object as a \gm-NLSy1 candidate to remain aligned with the classification definition of a NLSy1 source \citep[][]{1985ApJ...297..166O,1989ApJ...342..224G}. We perform the LAT data analysis for the remaining 252 sources by adopting the procedure described below. We also update the \gm-ray spectral parameters of the \gm-NLSy1s present in the 3FGL catalog following the same analysis steps.

We consider the Pass 8 source class photons covering the energy range of 100 MeV to 300 GeV and the first 101 months (2008 August 5 to 2017 January 5) of the \fermi~operation. We follow the standard data reduction steps as outlined in the online documentation\footnote{http://fermi.gsfc.nasa.gov/ssc/data/analysis/documentation/} and briefly described it here. We define a circular region of interest (ROI) of 15$^{\circ}$ radius centering on the NLSy1 galaxy and consider all sources lying within the ROI and included in the third \fermi-LAT catalog \citep[3FGL;][]{2015ApJS..218...23A}. Our sky model also consists of the Galactic and the isotropic extragalactic diffuse emission templates \citep[][]{2016ApJS..223...26A}. We perform a component-wise data analysis for all four point-spread function (PSF) type events. These four PSFs characterize the photons into four quartiles. The lowest quartile, PSF0, and the highest quartile, i.e., PSF3, represent the worst and the best, respectively, quality directional angular reconstruction.  By adopting a SUMMED likelihood method included in the pylikelihood library of the Science Tools\footnote{http://fermi.gsfc.nasa.gov/ssc/data/analysis/software/}, we optimize the spectral parameters of all the sources present in the model, including power-law normalization factors of the background diffuse models, so that the sky model represents the data as best as possible. Since the target objects are not present in the 3FGL, we model their spectral shapes as a simple power law and allow the prefactor and the photon index to vary during the likelihood optimization. The significance of the detection is computed in the form of the maximum likelihood test statistic TS=  2$\Delta \log \zeta$, where $\zeta$ represents the likelihood function, between models with and without a \gm-ray point source at the position of the NLSy1 galaxy. A source is considered to be detected if TS$>25$ \citep[4.2$\sigma$;][]{1996ApJ...461..396M}.

The 3FGL catalog represents the \gm-ray sky as observed by the \fermi-LAT in its first four years of the operation. In this work, we consider the LAT data covering more than 8 years. Therefore, it is possible that, similar to our target sources, there could be faint \gm-ray objects present in the data but not in the 3FGL catalog. To account for such unmodeled sources, we adopt an iterative procedure. We generate a residual TS map for the ROI and search for unmodeled excesses (with TS$>25$). Once found, their spatial positions are re-optimized and then inserted into the sky model. This procedure is repeated until the TS map stops showing excess emission.

\section{Results and Discussion}\label{sec:results}
\begin{figure*}
\hbox{\includegraphics[scale=0.6]{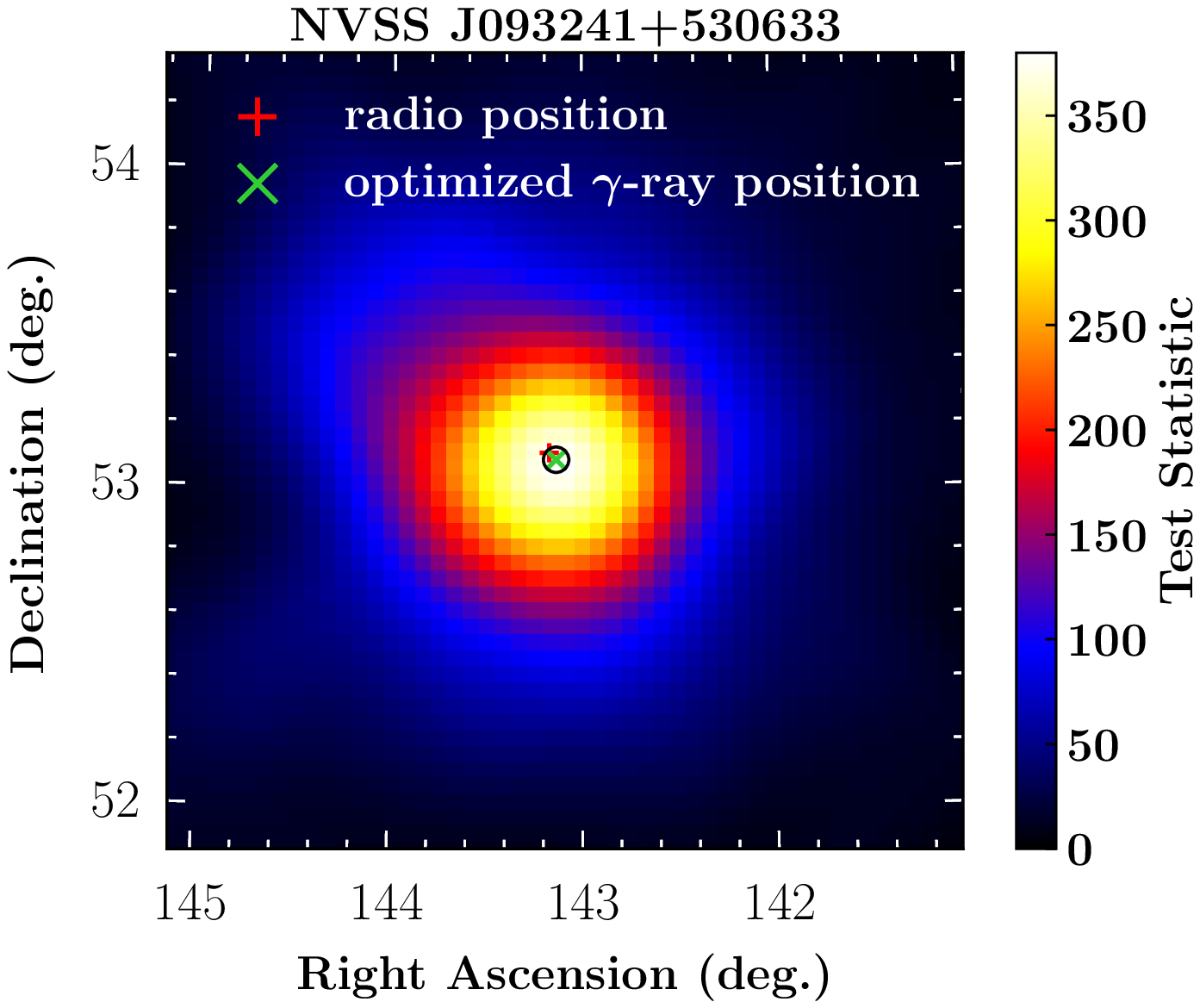}
          \includegraphics[scale=0.6]{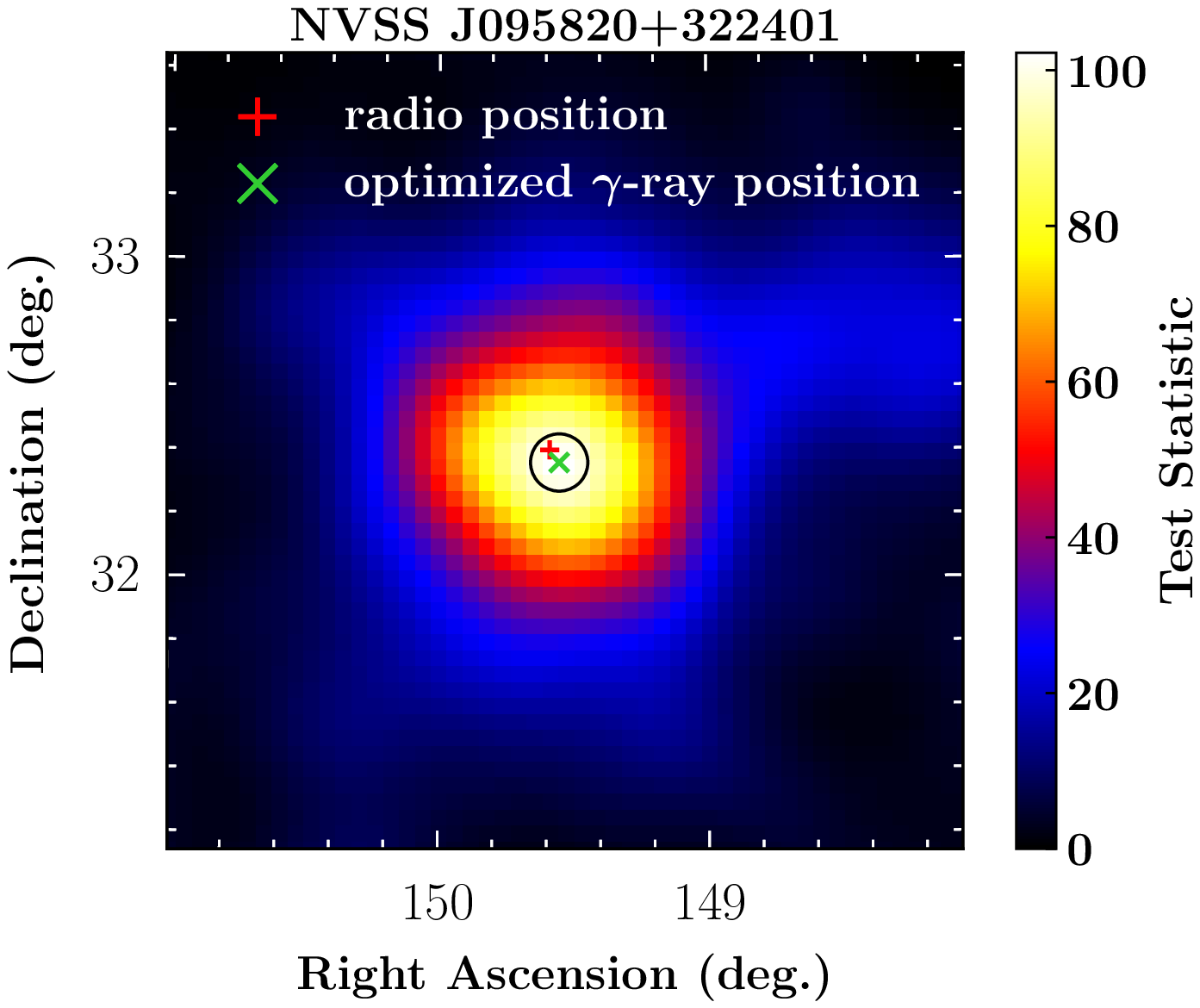}
            }
\hbox{\includegraphics[scale=0.6]{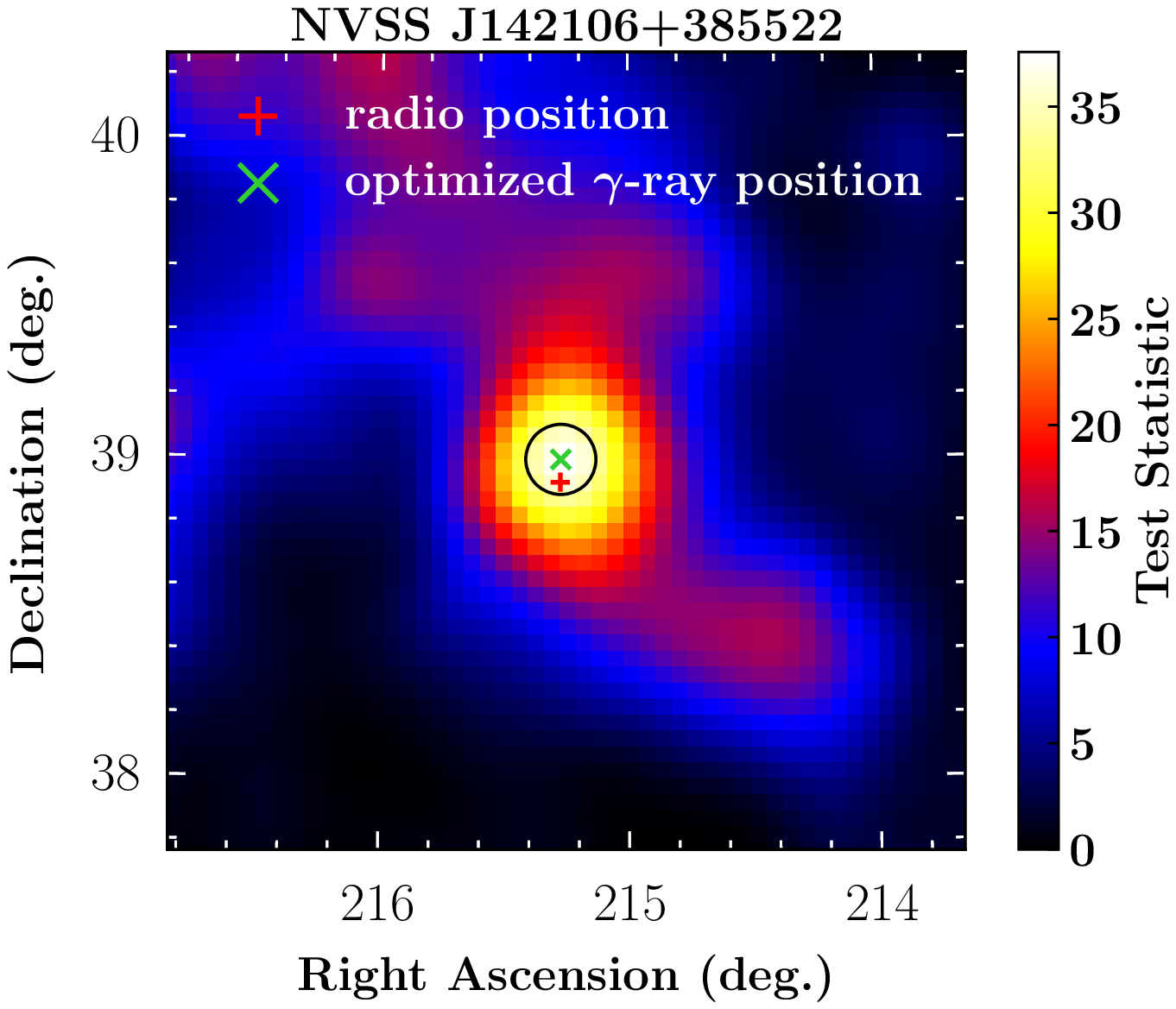}
         \includegraphics[scale=0.6]{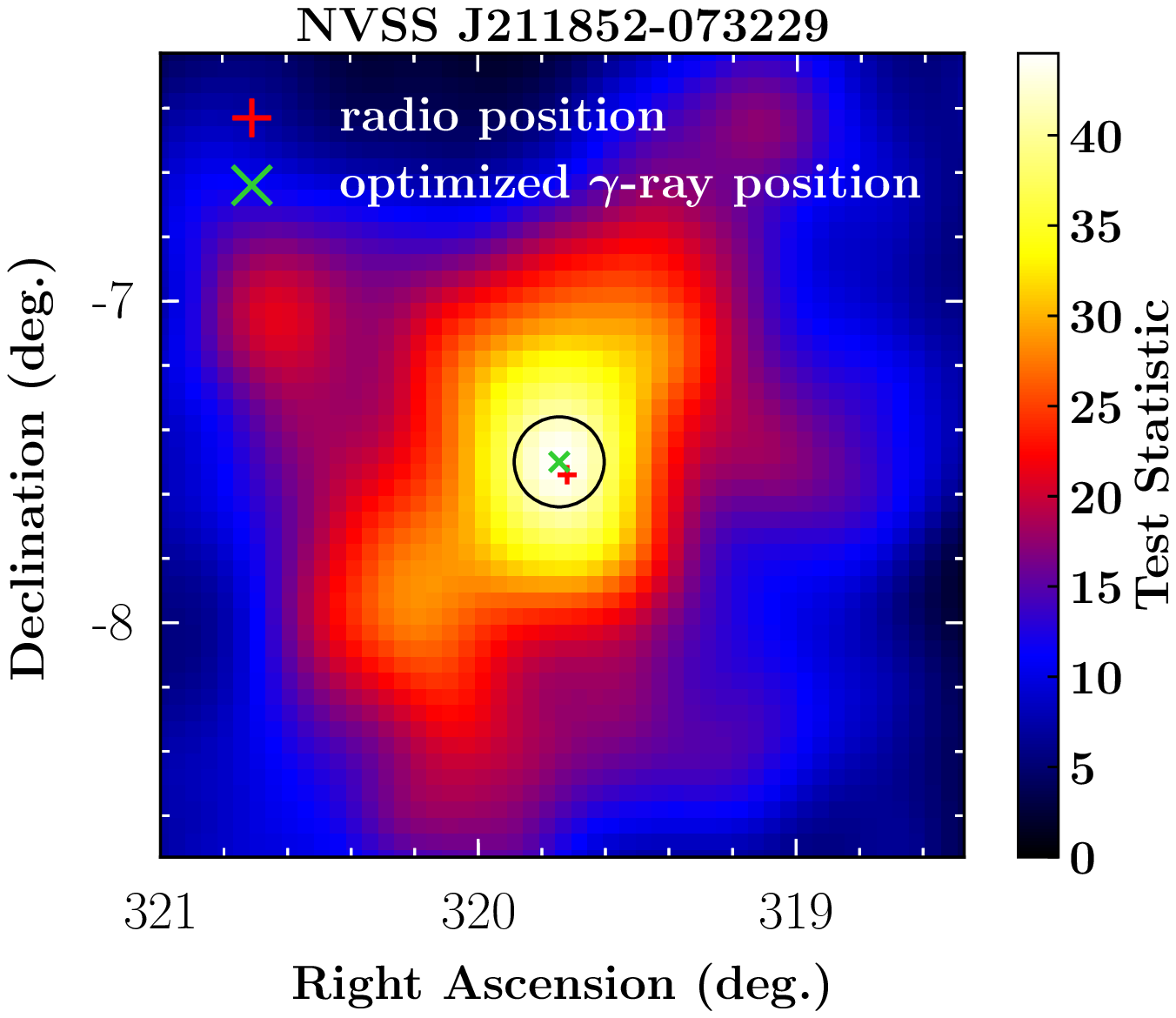}
}          
\caption{The test statistic maps of the four NLSy1 galaxies. The radio position (J2000), optimized \gm-ray position (J2000) are shown with red plus and green cross, respectively. The black circle denote the associated 95\% error circle.\label{fig:tsmap}}
\end{figure*}

A systematic search for \gm-ray emitting NLSy1s among the largest sample of NLSy1 sources, following the methodologies prescribed in the previous section, has led to the first \gm-ray detection of four new sources. To identify the association of the optimized \gm-ray positions with the NVSS counterparts, we perform the likelihood ratio association test \citep[see,][]{2011ApJ...743..171A} which confirms the NLSy1 galaxies to be the radio counterparts of the \gm-ray sources with high confidence (association probability $>$80\%). In Table \ref{tab:basic_info}, we provide the information of the newly detected \gm-NLSy1 galaxies and also present the results of the LAT data analysis\footnote{One of the newly \gm-ray detected source NVSS J142106+385522 ($z=0.49$) has incompleteness in its H$_{\beta}$ emission line profile, leading to the ambiguity in the FWHM measurement. \citet[][]{2011ApJS..194...45S} have reported it as 2935$\pm$1084 km s$^{-1}$, whereas, \citet[][]{2017ApJS..229...39R} derived it as 1616$\pm$186 km s$^{-1}$. Therefore, we include it in our final sample considering it as a candidate \gm-NLSy1 galaxy.}. The residual TS maps of the four \gm-ray detected NLSy1 galaxies are shown in Figure \ref{fig:tsmap} and we also show the radio and the optimized \gm-ray positions (in J2000). Furthermore, in Table \ref{tab:gamma_par}, we provide the updated spectral parameters for the NLSy1 sources present in the 3FGL catalog. 

In addition to the four new \gm-NLSy1s reported for the first time, three known \gm-ray emitting FSRQs are also found as NLSy1 sources by cross-matching the SDSS-DR12 NLSy1 catalog with 3LAC. The addition of these seven new objects has now significantly increased the sample size of the known \gm-NLSy1 galaxies. Similar to the previously known sources, the new \gm-NLSy1s exhibit a bright and compact radio core and a flat/inverted radio spectrum, providing another supporting evidence for the presence of a closely aligned relativistic jet.

\begin{figure*}
\hbox{
\includegraphics[scale=0.7]{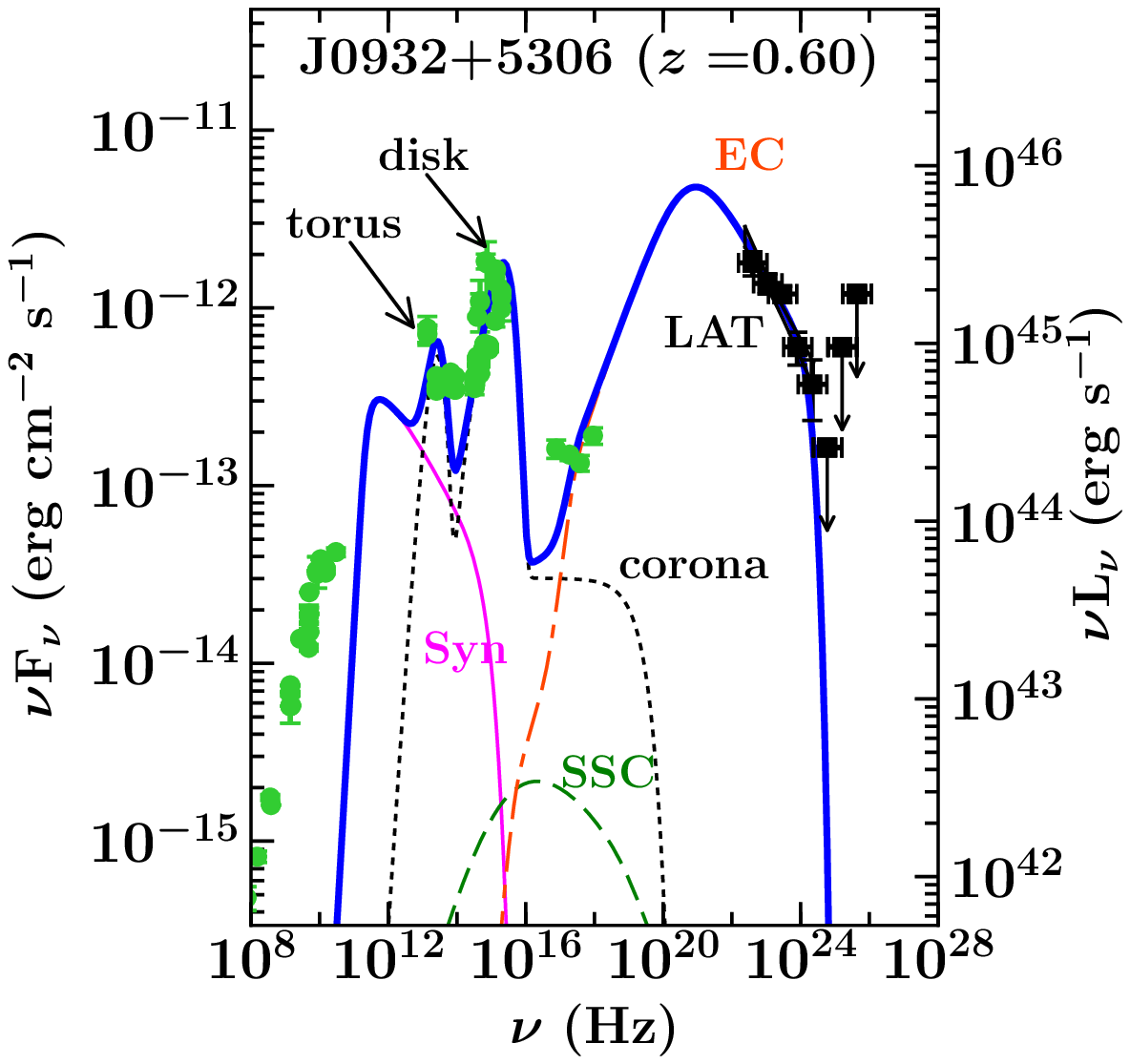}
\includegraphics[scale=0.7]{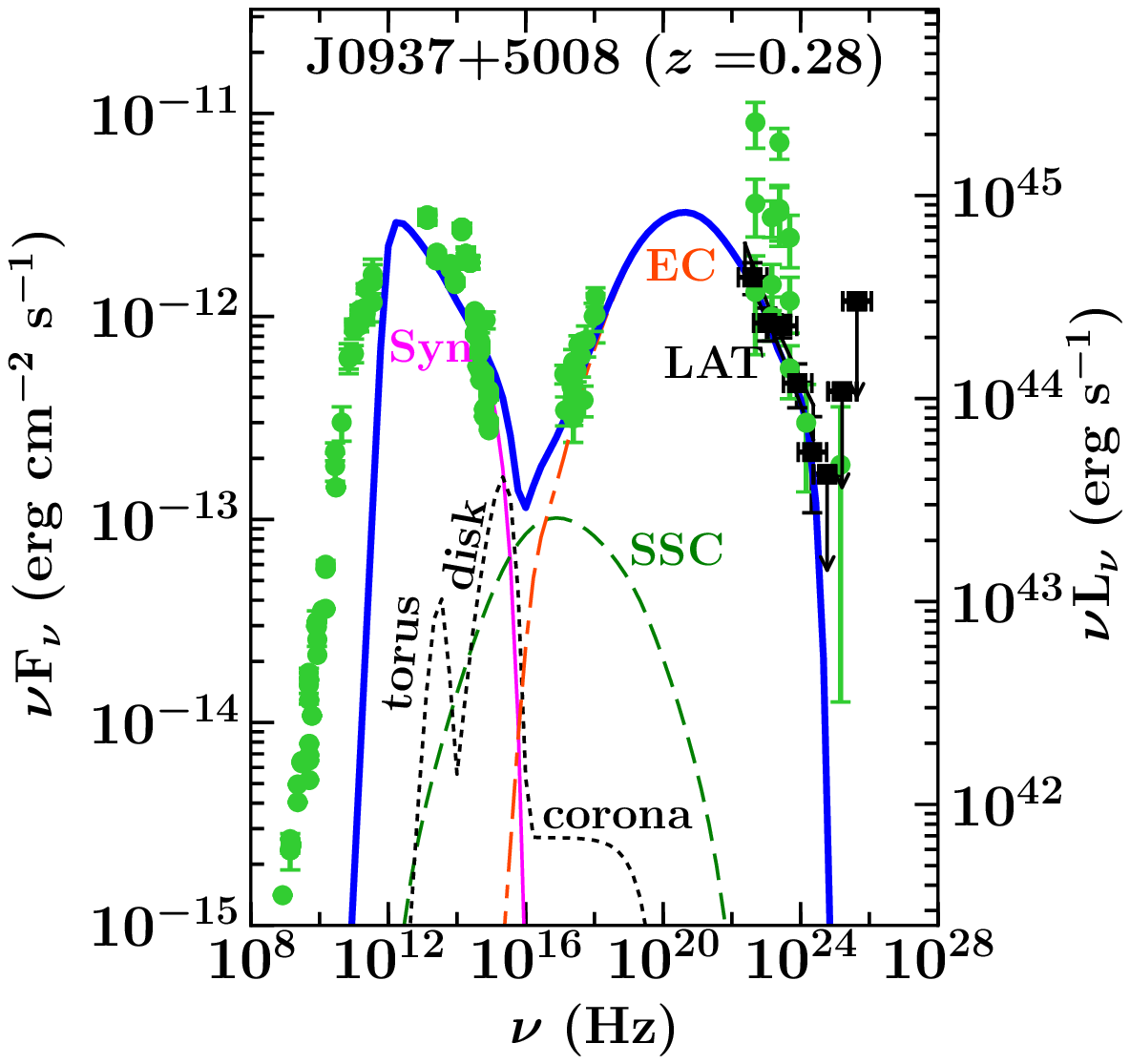}
}
\hbox{
\includegraphics[scale=0.5]{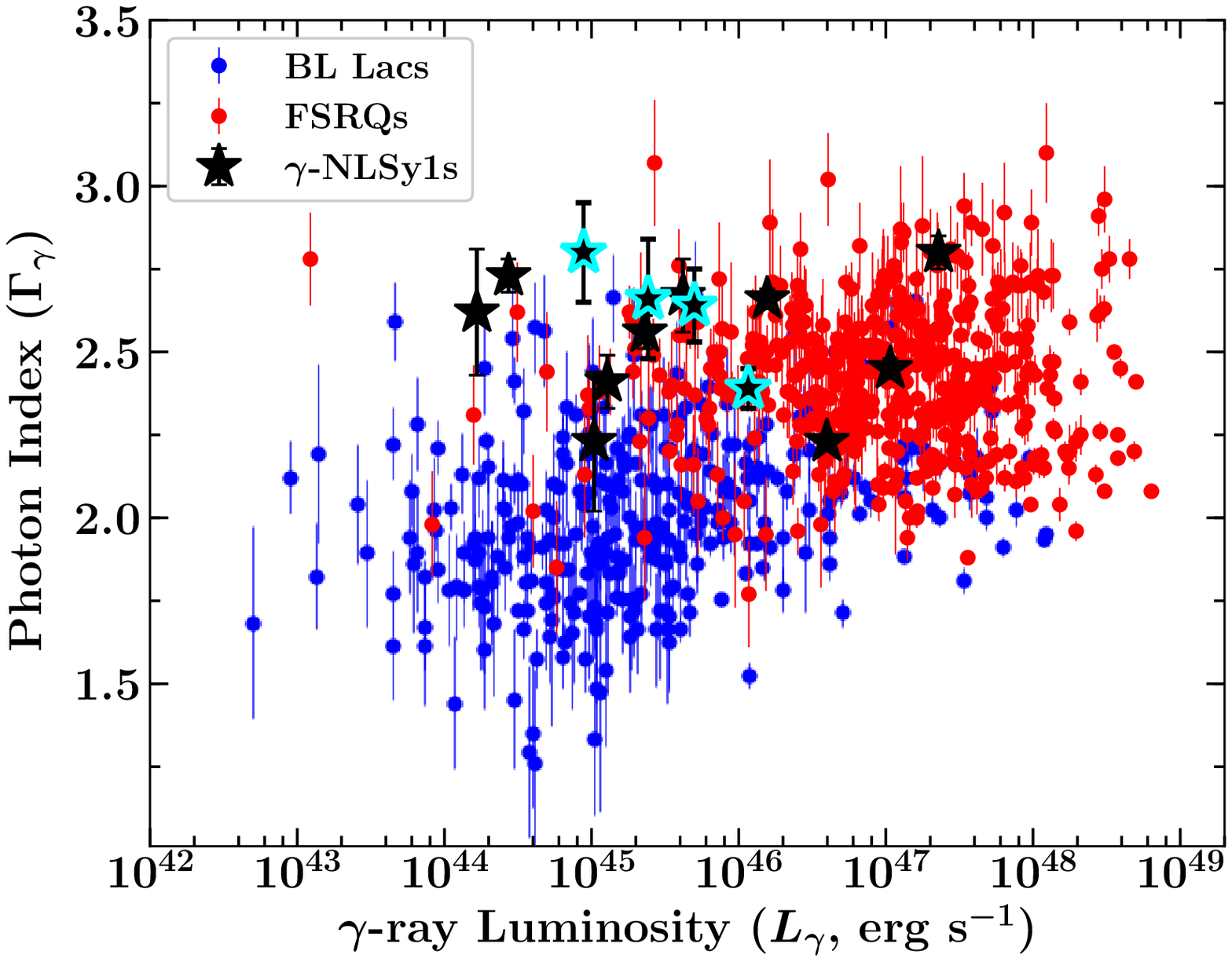}
\includegraphics[scale=0.5]{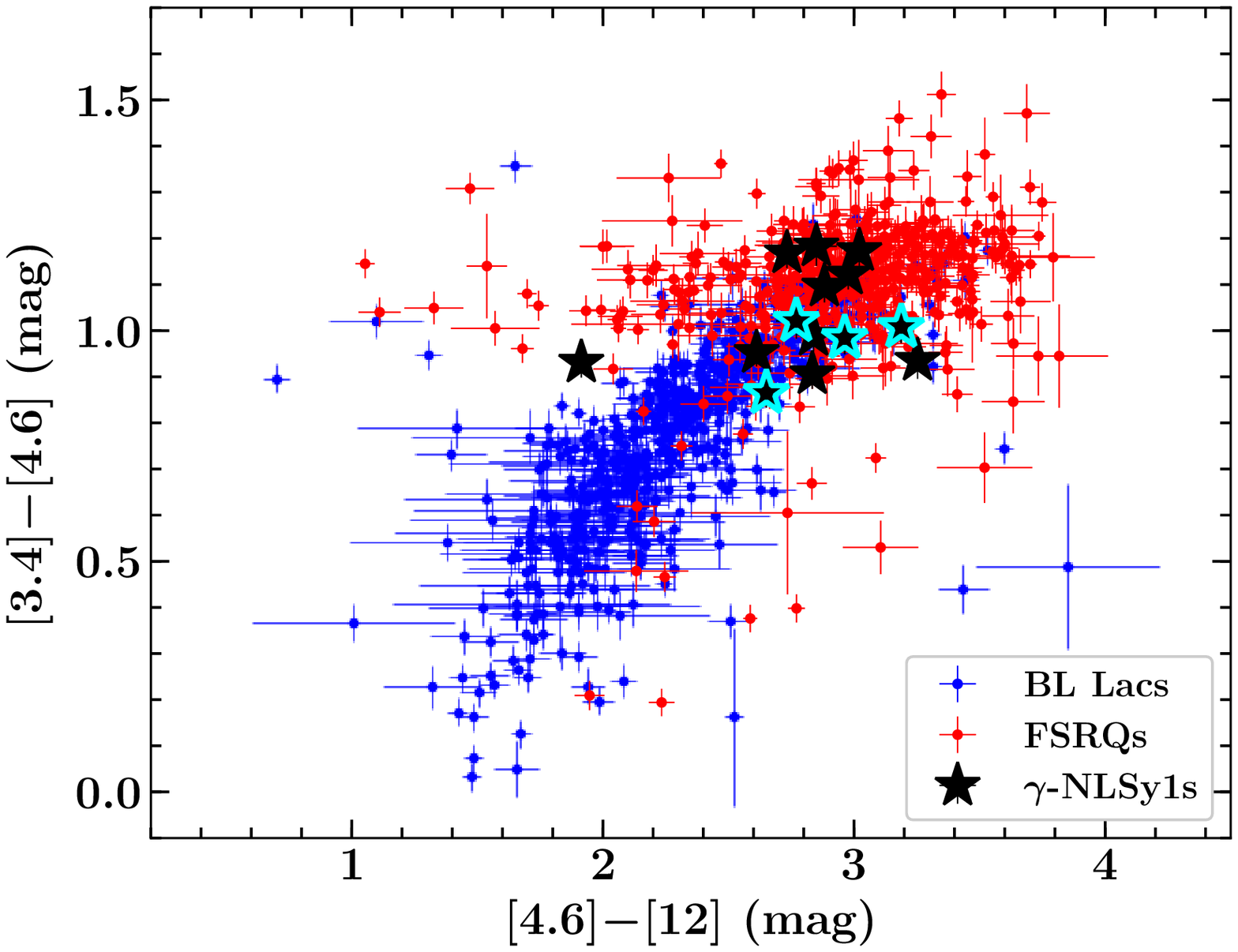}
}
\hbox{\hspace{4.0cm}
\includegraphics[scale=0.5]{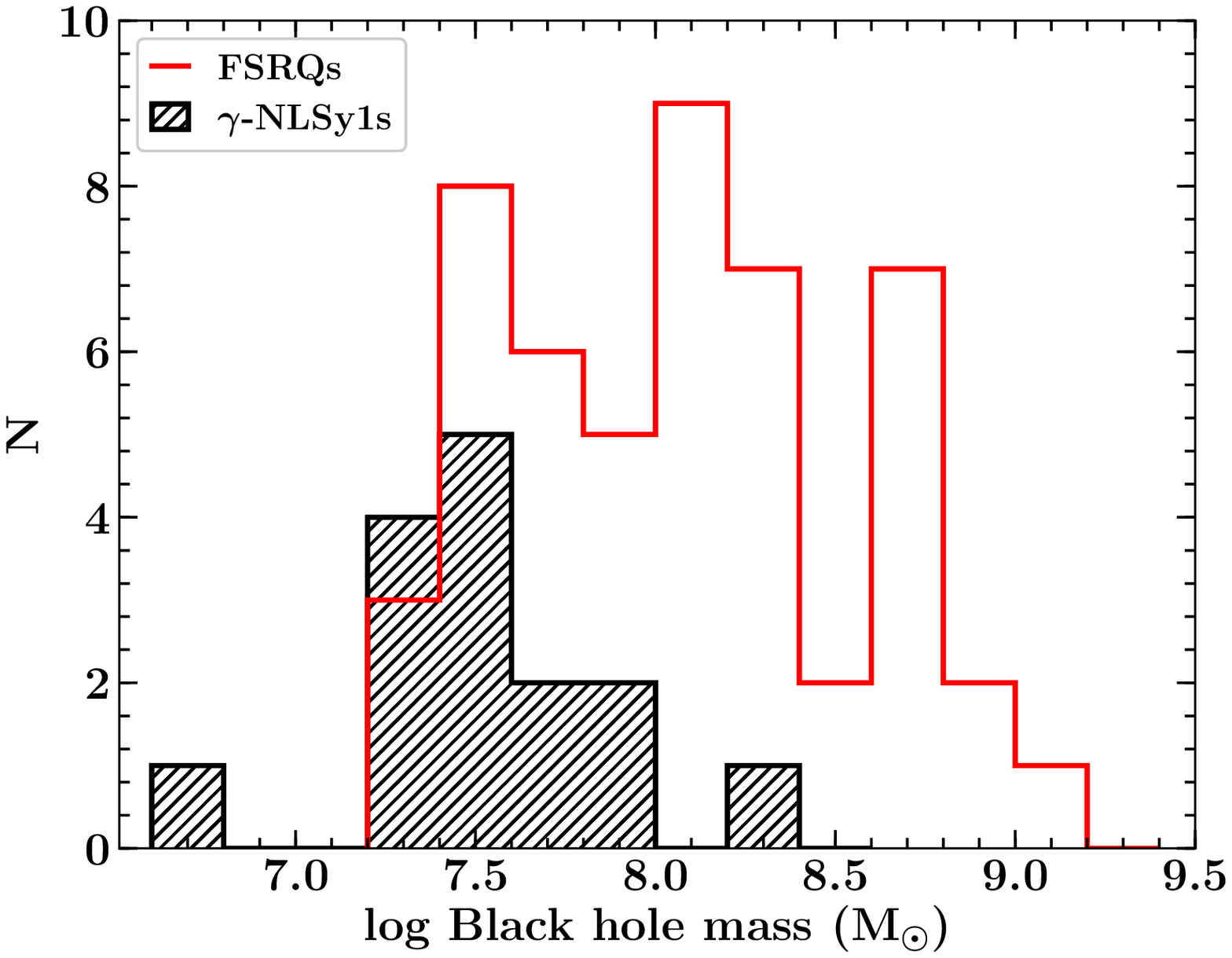}
}
\caption{Top: The broadband modeled SEDs of the \gm-NLSy1 galaxies. The lime green circles denote the archival observations, whereas, the black squares represent the \gm-ray spectrum derived from the LAT analysis. Downward arrows correspond to the 2$\sigma$ upper limits. Blue thick line is the sum of all the radiative components which are labeled. Middle left: Comparison of the \gm-ray luminosity and photon index values for all the known \gm-NLSy1 galaxies, including those identified in this work (black stars with cyan boundaries), with that for 3LAC blazars. Middle right: The WISE color-color diagram for the \fermi-LAT detected blazars. The \gm-NLSy1 galaxies are shown with black stars. Bottom: The distribution of the black hole mass derived from the H$_{\beta}$ emission line parameters for LAT detected FSRQs (red solid line) and \gm-NLSy1 galaxies (hatched black). See the text for details.\label{fig:3lac}}
\end{figure*}

We generate the broadband spectral energy distributions (SEDs) of two sources, NVSS J093241+530633 and GB6 J0937+5008, for which we could find the archival multi-wavelength data. The SEDs are modeled using a simple one-zone leptonic emission model \citep[see, e.g.,][for details]{2009MNRAS.397..985G}
and the results are shown in the top panel of Figure \ref{fig:3lac}.
The associated parameters for NVSS J093241+530633 are
as follows: broken power law spectral indices before ($p$) and after ($q$) the break= 1.8 and 3.7, respectively; magnetic field ($B$)= 1.3 Gauss; bulk Lorentz factor ($\Gamma$)= 10; break ($\gamma_{\rm b}$) and maximum ($\gamma_{\rm max}$) energies of the electron energy distribution= 53 and 3000, respectively; distance of the emission region ($R_{\rm diss}$)= 0.07 pc; size of the broad line region ($R_{\rm BLR}$)= 0.06 pc, disk luminosity ($L_{\rm disk}$)= 3.5$\times$10$^{45}$\lum. For 3FGL J0937.7+5008, we adopt the following SED parameters:  $p$=1.9, $q$=3.6, $B$= 1.7 Gauss; $\Gamma$= 17; $\gamma_{\rm b}$= 46; $\gamma_{\rm max}$= 3000; $R_{\rm diss}$= 0.01 pc; $R_{\rm BLR}$= 0.007 pc, $L_{\rm disk}$= 5.1$\times$10$^{43}$\lum. As can be seen, the sources exhibit a SED similar to blazars. Interestingly, the LAT spectrum is well explained by the external Compton process with the BLR and the dusty torus being the primary sources of the seed photons for the inverse Compton scattering. This indicates a stronger similarity with FSRQs than BL Lac type objects \citep[see, e.g.,][for similar results]{2012A&A...548A.106F,2013ApJ...768...52P}.

In the middle left panel of Figure \ref{fig:3lac}, we show the behavior of all of the \gm-NLSy1 galaxies on the \gm-ray photon index ($\Gamma_{\gamma}$) versus \gm-ray luminosity ($L_{\gamma}$) plane. Clearly, \gm-NLSy1 sources occupy a region where they have steep \gm-ray spectrum ($\Gamma_{\gamma}>2$), similar to FSRQs, and $L_{\gamma}$ smaller than powerful FSRQs. In other words, the \gm-ray properties of these objects suggest them to be low luminosity FSRQs \citep[see also,][for earlier results]{2009ApJ...699..976A}. These findings are in line with the predictions of \citet[][]{2009MNRAS.396L.105G} about the detection of the low black hole mass, low luminosity FSRQs with \fermi-LAT. The similarity of \gm-NLSy1 galaxies with FSRQs can also be seen in the Wide-field Infrared Survey Explorer (WISE) color-color diagram (middle right panel of Figure \ref{fig:3lac}). It has been noticed that the \gm-ray emitting blazars occupy a distinct place in this plot \citep[so called WISE blazar strip;][]{2011ApJ...740L..48M} and one can see that, on average, \gm-NLSy1 sources lie in a region mainly populated by FSRQs \citep[see also, a relevant discussion in][]{2015A&A...575A..13F}. Furthermore, rest-frame equivalent width of H$_{\beta}$ line for all of the \gm-NLSy1 galaxies are found to be larger than 5\AA~\citep[][]{2017ApJS..229...39R}, thus providing another supportive evidence of their similarity with FSRQs. 

We calculate or collect the $M_{\rm BH}$ values for \gm-NLSy1 galaxies using the H$_{\beta}$ line parameters reported in literature \citep[][]{2001ApJ...558..578O,2008ApJ...685..801Y,2017ApJS..229...39R} and show it in the bottom panel of Figure \ref{fig:3lac} (black hatched). For an equal comparison, we also show the $M_{\rm BH}$ distribution (red) for $z<1$ \gm-ray emitting FSRQs\footnote{Since all \gm-NLSy1 galaxies are nearby objects ($z<1$), we consider only FSRQs that have redshifts less than unity.} derived from the H$_{\beta}$ line information by \citet[][]{2012ApJ...748...49S}. It can be seen that \gm-NLSy1 galaxies 
occupy the low $M_{\rm BH}$ end of the FSRQ distribution \citep[see also,][]{2017FrASS...4....6F}. It should be noted that the black hole mass measurement in RL-NLSy1 galaxies is subject to considerable uncertainty and it was proposed that it is underestimated \citep[e.g.,][]{2008ApJ...678..693M,2017MNRAS.469L..11D}. However, since we use the same criteria (the cut on $z$ and considering only H$_{\beta}$ line parameters) to determine $M_{\rm BH}$ for both \gm-NLSy1s and FSRQs, a comparison between these two classes should be valid. Therefore, by comparing various observational properties of \gm-NLSy1 galaxies with blazars, as discussed above, it can be concluded that these sources are most likely the low black hole mass counterpart of powerful FSRQs and once normalized for $M_{\rm BH}$, these objects have physical properties similar to blazars \citep[][]{2015A&A...575A..13F}.

\begin{figure*}
\hspace{1.0cm}\includegraphics[width=\columnwidth]{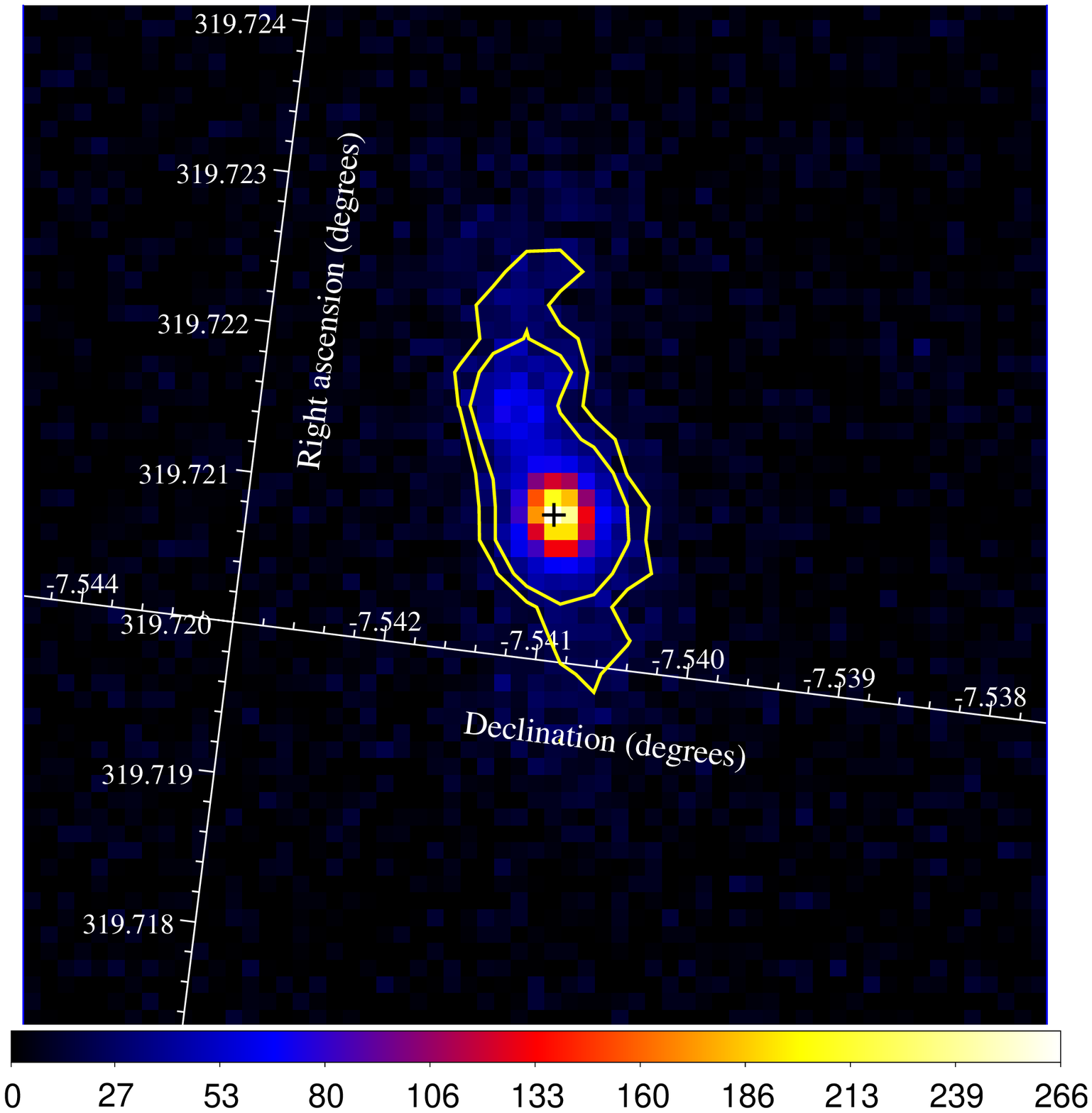}
\caption{SDSS $r'$ band image of NVSS J211852$-$073229 ($z=0.26$). The colorbar represents the SDSS count units. Confidence contours are at 3$\sigma$ and 5$\sigma$ levels and `+' mark shows the optical position of the source. An extended structure is evident.\label{fig:host}}
\end{figure*}

NLSy1s are typically hosted in spiral galaxies \citep[][]{2006AJ....132..321D} and the very few observations of \gm-NLSy1 galaxies also point to their spiral nature or recent mergers \citep[e.g.,][]{2016ApJ...832..157K,2017MNRAS.467.3712O}, though there are claims for the elliptical host also \citep[][]{2017MNRAS.469L..11D}. Interestingly, in the low resolution SDSS image of one of our sources, NVSS J211852$-$073229, we find suggestive evidences of an extended structure (see Figure \ref{fig:host}). Such an extended structure is often observed in barred spiral Seyfert galaxies \citep[e.g., NGC 7479; ][]{1960ApJ...132..654B} and these are claimed as the sites of intense star forming activities \citep[e.g.,][]{2011AJ....142...38Z}. Future high resolution imaging of these peculiar objects will throw more light on their host galaxy properties and resemblance/differences with blazars.

\begin{splitdeluxetable*}{lcccccccccBccccccc} 
\tabletypesize{\small}
\tablecaption{Information and results of the \fermi-LAT data analysis of newly \gm-ray detected NLSy1 galaxies.\label{tab:basic_info}}
\tablewidth{0pt}
\tablehead{
\colhead{} & \colhead{} & \colhead{} & \colhead{} & \colhead{Basic information} & \colhead{} & \colhead{}  & \colhead{} & \multicolumn{2}{c}{} & \colhead{} & \colhead{} & \colhead{} & \colhead{\fermi-LAT data analysis} & \colhead{} & \colhead{}\\
\colhead{Name} & \multicolumn{2}{c}{Radio position (J2000)} & \colhead{$F_{\rm 1.4~GHz}$} & \colhead{$g'$} & \colhead{$z$} & \colhead{FWHM}  & \colhead{[O~{\sc iii}]/H$_\beta$} & \colhead{$R_{\rm 4570}$} & \colhead{$M_{\rm BH}$} & \multicolumn{2}{c}{Optimized position (J2000)} & \colhead{$R_{95\%}$} & \colhead{$F_{\rm 0.1-300~GeV}$} & \colhead{$\Gamma_{\gamma}$} & \colhead{$L_{\gamma}$} & \colhead{TS}\\
\cline{2-3}
\cline{11-12}
\colhead{(NVSS)} & \colhead{hh mm ss.ss} & \colhead{dd mm ss.s} & \colhead{(mJy)} & \colhead{(mag)} & \colhead{} & \colhead{(km s$^{-1}$)} & \colhead{} & \colhead{} & \colhead{$M_{\sun}$} & \colhead{hh mm ss.ss} & \colhead{dd mm ss.s} & \colhead{(degrees)} & \colhead{(10$^{-9}$ \phflux)} & \colhead{} & \colhead{(10$^{45}$ \lum)} & \colhead{}
}
\startdata
J093241+530633	  & 09 32 41.1  & +53 06 33.3    & 482  & 18.9 & 0.60 & 1897$\pm$181 &  1.46 & 0.27 & 7.66 & 09 32 32.5  & +53 05 15.2    & 0.04 & 11.3$\pm$1.2 & 2.39$\pm$0.06 & 11.6$\pm$1.8  & 364\\
J095820+322401    & 09 58 20.9	  & +32 24 01.6   &1247 & 16.0 & 0.53 & 976$\pm$48 & 2.88 & 0.66    & 7.72 & 09 58 12.5  & +32 21 38.8    & 0.09 & 7.9$\pm$1.4   & 2.64$\pm$0.11 & 5.0$\pm$1.2    & 97\\
J142106+385522    & 14 21 06.0	  & +38 55 22.5   & 85.7 & 18.6 & 0.49 & 1616$\pm$186 & 0.95 & 0.45    & 7.36 & 14 21 05.5  & +38 59 39.6    & 0.11 & 4.8$\pm$1.7   & 2.66$\pm$0.18 & 2.4$\pm$1.1    & 36\\
J211852$-$073229 & 21 18 52.9  & $-$07 32 29.3 & 96    & 16.5 & 0.26 & 1833$\pm$243 & 1.63 &  0.26 & 7.21 & 21 18 58.9  & $-$07 30 04.5 & 0.14 & 9.2$\pm$2.0   & 2.80$\pm$0.15 & 0.9$\pm$0.3    & 40\\
\enddata
\tablecomments{Name, radio positions, and 1.4 GHz flux values have been taken from the NVSS catalog. The $g'$ band magnitude and redshifts are adopted from SDSS. $R_{\rm 4570}$ is the optical Fe~{\sc ii} strength relative to broad component of the H$_{\beta}$ emission line. $M_{\rm BH}$ is the logarithmic central black hole mass, in units of solar mass, derived from optical spectroscopic emission line parameters reported in \citet[][]{2017ApJS..229...39R} and following the empirical relation of \citet[][]{2011ApJS..194...45S}. $R_{95\%}$ is the 95\% error radius derived from the \gm-ray analysis. The \gm-ray flux and apparent luminosity are in the energy range of 0.1$-$300 GeV.}
\end{splitdeluxetable*}

\begin{table*}
\begin{center}
\caption{Spectral parameters derived from the LAT data analysis of the known NLSy1 objects present in the 3FGL catalog.\label{tab:gamma_par}}
\begin{tabular}{cccccccc}
\tableline
Name & Counterpart & Model & $F_{\gamma}$ & $\Gamma_{\gamma}$/$\alpha$ & $\beta$ & TS & $L_{\gamma}$ \\
~[1] & [2] & [3] & [4] & [5] & [6] & [7] & [8]\\
\tableline
3FGL J0325.2+3410           & 1H 0323+342                        & LP & 6.76$\pm$0.26   & 2.73$\pm$0.05 & 0.10$\pm$0.03 & 1531 & 0.26$\pm$0.02 \\
3FGL J0849.9+5108           & SBS 0846+513                      & PL & 3.35$\pm$0.13   & 2.23$\pm$0.02 & ---                      & 2890 & 39.83$\pm$2.48 \\
3FGL J0937.7+5008$^*$   & GB6 J0937+5008                   & PL & 0.86$\pm$0.12   & 2.41$\pm$0.08 & ---                     & 221   & 1.28$\pm$0.27 \\
3FGL J0948.8+0021           & PMN J0948+0022                  & LP & 11.70$\pm$0.26 & 2.45$\pm$0.03 & 0.14$\pm$0.02 & 6328 & 108.20$\pm$4.56 \\
3FGL J1222.4+0414           & CGRaBS J1222+0413             & LP & 7.25$\pm$0.30   & 2.80$\pm$0.05 & 0.05$\pm$0.03 & 1843  & 227.60$\pm$11.77 \\
3FGL J1505.1+0326           & PKS 1502+036                      & PL & 4.90$\pm$0.22   & 2.66$\pm$0.03 & ---                      & 1327 & 15.60$\pm$0.93 \\
3FGL J1520.3+4209$^*$    & TXS 1518+423                     & PL & 0.85$\pm$0.15  & 2.67$\pm$0.11 & ---                       & 100   & 4.18$\pm$0.96 \\
3FGL J1644.4+2632           & RGB J1644+263                    & PL & 1.28$\pm$0.18   & 2.76$\pm$0.10 & ---                      & 117 & 0.32$\pm$0.06 \\
3FGL J2007.8$-$4429        & PKS 2004$-$447                  & PL & 2.50$\pm$0.18  & 2.56$\pm$0.05 & ---                       & 534   & 2.30$\pm$0.24 \\
3FGL J2118.4+0013$^*$    & PMN J2118+0013                 & PL & 0.30$\pm$0.10  & 2.23$\pm$0.21 & ---                       & 31     & 2.01$\pm$1.20 \\
\tableline
\end{tabular}
\end{center}
\tablecomments{The column information are as follows: col.[1]: name of the source. Names with asterisk are the sources which are reported as new \gm-NLSy1s in this work, including GB6 J0937+5008 a candidate \gm-NLSy1; col.[2]: counterpart name; col.[3]: spectral model reported in the 3FGL catalog, PL: power law and LP: log parabola; col.[4]: 0.1$-$300 GeV \gm-ray flux, in units of 10$^{-8}$ \phflux; col.[5]: slope of the power law model or slope at pivot energy in the log parabola model; col.[6]: curvature parameter in the log parabola model, col.[7]: test statistic; and col.[8]: \gm-ray luminosity in the energy range of 0.1$-$300 GeV, in units of 10$^{45}$ \lum.}
\end{table*}

\acknowledgments
The \textit{Fermi}-LAT Collaboration acknowledges support for LAT development, operation and data analysis from NASA and DOE (United States), CEA/Irfu and IN2P3/CNRS (France), ASI and INFN (Italy), MEXT, KEK, and JAXA (Japan), and the K.A.~Wallenberg Foundation, the Swedish Research Council and the National Space Board (Sweden). Science analysis support in the operations phase from INAF (Italy) and CNES (France) is also gratefully acknowledged. We acknowledge the use of the SDSS and WISE data. S.R. acknowledges the support by the Basic Science Research Program through the National Research Foundation of Korea government (2016R1A2B3011457).
\vspace{5mm}
\facilities{\fermi-LAT}


\end{document}